\documentstyle[12pt,aasms4]{article}

\lefthead{Tsuji }
\righthead{Water in emission in the early M supergiant stars mu Cephei}

\begin{document}

\title{
WATER IN EMISSION IN THE ISO SPECTRUM OF THE EARLY M SUPERGIANT 
STAR  $\mu$ CEPHEI\footnote{Based on the Data Archives of ISO, an ESA
project with instruments funded by ESA Member States (especially the PI
countries: France, Germany, the Netherlands and the United Kingdom) and
with the participation of ISAS and NASA.} 
}

\author{Takashi Tsuji}
\affil{Institute of Astronomy, The University of Tokyo,
 Mitaka, Tokyo, 181-0015 Japan}

\begin{abstract}

We report a detection of water in emission in the spectrum of the
M2 supergiant star  $\mu$ Cep (M2Ia) observed by the 
Short Wavelength Spectrometer (SWS) aboard  Infrared Space Observatory
(ISO) and now released as the ISO Archives. The emission first
appears in the 6 $\mu$m region ($\nu_{2}$ fundamental bands), and
then in the 40 $\mu$m region (pure rotation lines) despite
the rather strong dust emission. The
intensity ratios of the emission features are far from those of
the optically thin gaseous emission. Instead, we could reproduce the major
observed emission features by an optically thick  water sphere of the
inner radius about two stellar radii ($2 R_{*} \approx 
1300 R_{\odot}$), $T_{\rm ex} = 1500$K, and $N_{\rm col}$(H$_2$O)$ = 
3 \times 10^{20}$ cm$^{-2}$. This model also accounts for the  H$_{2}$O
absorption bands in the near infrared (1.4, 1.9, and 2.7$\mu$m) 
as well. The detection of water in emission provides strong constraints
on the nature of water in the early M supergiant star,
and especially its origin in the outer atmosphere is confirmed
against other models such as the large convective cell model.  
We finally confirm that the early M supergiant star
is surrounded by a huge optically thick sphere of the warm water vapor,  
which may be referred to as MOLsphere for simplicity.
Thus, the outer atmosphere of M supergiant stars should have
a complicated hierarchical and/or hybrid structure
with at least three major constituents including the warm MOLsphere
($T \approx 10^3$K) together with the previously known
hot chromosphere ($T \approx 10^4$K) and  cool expanding gas-dust 
envelope ($T \approx 10^2$K).  

\end{abstract}


\section{INTRODUCTION}

Presence of water in the stellar environment has been known since 1960's, 
especially by the pioneering space observations with the balloon-borne 
telescope Stratoscope II (Woolf, Schwarzschild, \& Rose 1964).
However, it is the recent  ISO mission (Kessler et al. 1996) that 
the infrared spectra unobscured by Earth's  atmosphere have finally been 
explored in full details. One of the important results of the highly 
successful ISO mission is that water exists everywhere in the universe
and that the star is not an exception. In fact, water has been detected
not only by absorption but also by emission in cool luminous  variables 
such as W Hya (Barlow et al. 1996; Neuffeld et al. 1996), NML Cyg 
(Justtanont et al. 1996), S Per (Tsuji et al. 1998), SW Vir (Tsuji, Aoki, 
\& Ohnaka 1999), R Cas (Truong-Bach et al. 1999), $o$ Cet (Yamamura, 
de Jong, \& Cami 1999), VY CMa (Neuffeld et al. 1999) etc. The exact nature 
of these emission features, however, is by no means clear yet.

On the other hand, water in the early  M (super)giant stars has been paid 
little attention until recently, since water has not been detected in these 
objects nor expected to be abundant in their atmospheres. Nevertheless, 
water in absorption has finally been detected  by ISO in the early M  
(super)giants (Tsuji et al. 1997, 1998) as well as by the ground based 
mid-infrared observation in the early M supergiant stars (Jennings \& Sada 
1998). More correctly, however, water in the early M supergiants had in fact 
been discovered in the spectra taken by Stratoscope II (Woolf et al. 1964; 
Danielson, Woolf, \& Gaustad 1965) more than 30 years before the ISO mission,
although this discovery has been overlooked until recently in favor of the 
opposing interpretation. We have shown that the Stratoscope observers had 
correctly identified water in the early M supergiant stars but that the 
water detected by them is not originating in the photosphere (Tsuji 2000). 

To clarify the nature of water found in the early M supergiant stars,
ISO should provide the best possibility. In fact, thanks to the recently 
released ISO Archives, we found water in emission in $\mu$ Cep for the 
first time in such an early M supergiant star. This detection is somewhat 
unexpected, but this provides not only the definitive evidence for the 
presence of water in the outer atmosphere of the early M supergiant stars 
but also a new clue to understanding the nature of water 
in the stellar environment.

\section{OBSERVED EMISSION LINES OF WATER}

We retrieved the spectra of $\mu$ Cep (M2Ia: observed on 1996 Dec.18 UT)
and  $\alpha$ Ori (M2Iab: observed on 1997 Oct.08 UT) from the ISO Archives.
These spectra were observed with the SWS (de Graauw et al. 1996) by its 
highest resolution grating mode (resolution $ R \approx 1600$ or about
188 km s$^{-1}$) in the region between 2.4 and 45  $\mu$m.
The region of the H$_{2}$O $\nu_{1}$ and $\nu_{3}$ 
fundamentals around 2.7 $\mu$m is somewhat complicated due to
the photospheric CO and OH bands, but the ISO data of both
$\alpha$ Ori and $\mu$ Cep show additional absorption features not 
predicted by the photospheric models. At least, part of these
features should be the same origin as the 1.4 and 1.9 $\mu$m absorption 
bands, namely H$_{2}$O of the non-photospheric origin (Tsuji 2000). 

\placefigure{fig1} 

The region of the H$_{2}$O $\nu_{2}$  fundamentals (the band origin at 
6.3 $\mu$m) is shown in Fig. 1. First, the spectrum of $\alpha$ Ori 
shows only weak absorption features due to  H$_{2}$O (indicated by the 
dotted lines in Fig.1a), as can be confirmed by the comparison with
the H$_{2}$O data based on HITEMP (Rothman 1997) shown in Fig.1c, but 
this detection is the best evidence for water in  $\alpha$ Ori at present.
The spectrum of $\mu$ Cep shown in Fig.1b is quite different
with complicated spectral features which show upturn at about 6.4 $\mu$m
in contrast to a smooth decline typical of the Rayleigh-Jeans spectrum as 
in $\alpha$ Ori. This upturn is  explained by the onset of emission due 
to the $P$ branch of the H$_{2}$O $\nu_{2}$ fundamentals.
Thus, the H$_{2}$O bands in absorption at 1.4, 1.9 and 2.7 $\mu$m turn
to emission at 6 $\mu$m, and this can be regarded as definite evidence 
for the presence of water in the extended outer atmosphere of $\mu$ Cep. 
Also, some emission features of the observed spectrum are well identified 
with the fine structures of H$_{2}$O bands by the comparison with the 
H$_{2}$O data (Fig.1c). The intensity ratios of the emission features 
do not agree with those of the optically thin emission represented by the 
absorption cross-section of H$_{2}$O at $T = 1500$K in Fig.1c, 
and the situation cannot be improved by changing the temperature. 

The spectrum of $\mu$ Cep 
has once been observed by the NASA Airborne Infrared Observatory
(Russell, Soifer, \& Forrest 1975), and non-grey behavior from
5 to 8 $\mu$m has been noted. The observers preferred to assume
excess absorption in the 4-5 $\mu$m rather than excess flux in 
the 5-8 $\mu$m. Another interpretation was to attribute the observed 
5-8 $\mu$m feature as due to emission of H$_{2}$O $\nu_{2}$ fundamentals, 
and  this was also proposing that water, besides dust, should be an 
important constituent of the outer atmosphere of $\mu$ Cep (Tsuji 1978). 
There has been no means by which to test such
a proposition at that time, but this is now fully confirmed by the ISO SWS 
which clearly resolved the fine structure of the H$_{2}$O emission bands. 

We already know that H$_{2}$O pure rotation lines were observed as
absorption in $\alpha$ Ori by the ground-based high resolution
spectroscopy in the 12 $\mu$m region (Jennings \& Sada 1998).
However, the spectral resolution of the ISO SWS is not high enough
to detect such faint lines. In the 40 $\mu$m region of $\alpha$ Ori, some 
emission-like features can be seen (Fig.2a), but they cannot be 
identified  with the known sources, except for an emission 
feature at 40.5 $\mu$m which roughly corresponds to  H$_{2}$O lines 
as can be known by the comparison with the H$_{2}$O data (Fig.2c).  
However, it is still possible that the observed features are due to 
the mixture of absorption (including photospheric lines) and emission 
unresolved by the low  resolution.

\placefigure{fig2} 

More distinct emission features appear in $\mu$ Cep (Fig.2b) and most of 
them  can be identified with the H$_{2}$O pure rotation lines by the 
comparison with the H$_{2}$O data (Fig.2c). The intensity  ratios of the 
emission features  again suggest that the emission should be originating 
from an optically thick source. Such emission features are relatively 
well seen in the 40 $\mu$m region, but not so clear in the other 
mid-infrared region.  One reason for this may be the presence of dust 
emission. To illustrate this, the entire spectrum of $\mu$ Cep 
observed by the ISO SWS is shown in Fig.3a by the red dots. It is clear 
that dust emission dominates the region longward of 8 $\mu$m.

\placefigure{fig3} 

\section{ORIGIN OF THE WATER EMISSION}

We showed that a  hypothetical absorption slab of H$_{2}$O with
$ T_{\rm ex} = 1500$K and $N_{\rm col} = 3.0 \times 10^{20}$ cm$^{-2}$ 
above the photosphere explains the 1.4 and 1.9 $\mu$m bands observed
in the Stratoscope spectrum of $\mu$ Cep (Tsuji 2000). The presence of
emission at the longer wavelength in the spectrum of $\mu$ Cep confirms 
that such a model of the non-photospheric origin of the water spectrum 
is basically correct, but we must  now introduce geometrical extension 
of the water gas to explain the formation of emission features.
For this purpose, we replace the simple absorption slab by
a spherically symmetric isothermal gaseous envelope of water,
whose inner and outer radii are $ R_{in} $ and $ R_{out} $, respectively. 
The density distribution in this envelope is assumed
to be in hydrostatic equilibrium under the gravity field of the central
star, and hence major contribution to the spectrum comes from the
relatively dense region near $ R_{in} $. For this reason, $ R_{in} $
is more important than $ R_{out} $. 

We already suggested that the emission features may be originating
from an optically thick source (Sect.2). In fact, the column density 
($N_{\rm col} = 3.0 \times 10^{20}$ cm$^{-2}$) and the absorption 
cross-section of H$_{2}$O typically $10^{-17}$ cm$^{2}$ mol$^{-1}$ at 
$T = 1500$K (Figs. 1c and 2c) suggest that the optical depth of the 
strong lines should be as large as $10^{3}$. Thus, the 
water envelope is optically quite thick in the lines, and the warm 
molecular envelope can be  referred to as molecular sphere, or MOLsphere 
for simplicity. Then, we must solve transfer equation in this MOLsphere 
and, since we are to interpret line spectra,
the photospheric spectrum including the full details of the absorption
lines  should exactly be considered as the boundary condition.
For this purpose, we apply the photospheric spectrum characterized by 
$M_{*}=15\,M_{\odot}, R_{*} = 650\,R_{\odot}$, $L_{*} \approx 6 \times 
10^{4}\,L_{\odot}$ and $T_{\rm eff} \approx 3600$\,K (Fig.1 of Tsuji 2000).
The computation has been done  with the spectral resolution of about 50000
by the use of the H$_{2}$O line database HITEMP (Rothman 1997). 
After a few trials, we found that $ R_{in} \approx 2  R_{*} \approx 
1300 R_{\odot} $  results in a spectrum that agrees with the observed 
one reasonably well. We also assumed $ R_{out} \approx 4 R_{*}$, but 
this is of minor importance by the reason outlined above. The resulting 
emergent spectrum  is shown by the green line in Fig.3a.

In the spectral region shortward of 5 $\mu$m, the MOLsphere
contributes only to  absorption, and the upper boundary of the 
spectrum shown by the green line in Fig.3a represents the photospheric 
continuum, below which CO, OH, and SiO bands of the
photospheric origin can be seen together with the H$_{2}$O
absorption produced anew in the MOLsphere (molecules other
than H$_{2}$O are not considered in the MOLsphere). At the shorter and
longer wavelength sides of 6.3 $\mu$m, the $R$ and $P$ branches,
respectively, of the H$_{2}$O $\nu_{2}$  fundamentals are seen in emission
above the photospheric continuum (which is not explicitly shown but
can be extrapolated from the near infrared continuum). This emission
of water due to the MOLsphere remains prominent throughout the mid
infrared region because of the stronger H$_{2}$O pure
rotation lines. The MOLsphere contributes only to emission in the 
mid-infrared region and the lower boundary of the spectrum shown by the 
green line represents the photospheric continuum, below 
which OH pure rotation lines of the photospheric origin can be seen. 
But most of these OH lines are weaker or some OH lines are missing as 
compared with those in the photospheic spectrum because of the filling 
in by the overlapping H$_{2}$O emission lines from the MOLsphere.
 
For comparing the predicted spectrum from the MOLsphere discussed above 
with observations, we convolve it with the
slit function of SWS which is assumed to be the Gaussian of 
FWHM = 187.5 km s$^{-1}$ and the resulting low resolution spectrum is 
shown by the blue line in Fig.3a. The result can directly be compared with 
the observed spectrum in the region shortward of 8 $\mu$m where
dust emission is negligible, and is shifted vertically to fit the observed 
spectrum of $\mu$ Cep shown by the red dots. A close-up of the region 
around 6.3 $\mu$m is shown in Fig.3b, where the red dots represent the
observed spectrum and the blue line the predicted spectrum from the 
MOLsphere, as in Fig.3a. The agreements between the observed
and predicted spectra appear to be fine if not perfect.

For interpreting the spectrum lonward of 8 $\mu$m, effect of dust emission 
on our predicted spectrum should be taken into account. Since 
dust envelope of $\mu$ Cep is optically thin (e.g. Tsuji 1978), 
we simply add the dust emission to the spectrum by the MOLsphere. 
The region near 40$\mu$m is shown by an expanded scale in Fig.3c, where 
the red dots represent the observed spectrum 
and the black line the predicted spectrum which is the
sum of the spectrum from the MOLsphere (the blue line in Fig.3a) and dust
emission represented by the Rayleigh-Jeans spectrum normalized to 162 Jy 
at 40 $\mu$m. The agreements between the observed
and predicted spectra appear to be reasonable.
This result also shows that the mid-infrared flux of apparently
dusty stars such as  $\mu$ Cep consists not only of dust emission but
also of gaseous emission of water.  The contribution of water
to the thermal balance in the outer atmosphere as a whole is not very large
(only 10 \% or so in the case of $\mu$ Cep), yet water may be the
dominant coolant in the inner part of the outer atmosphere.

In the above analysis, we apply  $ T_{\rm ex}$ and $N_{\rm col}$ for  
H$_{2}$O in the MOLsphere determined by the previous analysis of 
the Stratoscope spectra (Tsuji 2000). We confirmed that these parameters 
explain the ISO spectra as well and no improvement is  gained if these 
parameters are changed. Certainly, these parameters are better determined
from absorption rather than from emission spectra.
Also, the near infrared spectrum predicted from
the  MOLsphere agrees well with the Stratoscope spectrum, which was
interpreted  by the simpler slab model in our previous analysis. 
The new constraint imposed by the ISO spectra is essentially the
inner radius of the MOLsphere,  $ R_{in} $. In fact, 
the  emission features are quite sensitive to $ R_{in} $, and
the emission tends to be much stronger for  $ R_{in}$  larger than  
$ R_{in} \approx 2 R_{*}$ we have adopted. Further, 
the 6 $\mu$m emission turns to absorption for the smaller values, e.g.
$ R_{in} \approx 1000 R_{\odot} \approx 1.5 R_{*}$.  The weak 6 $\mu$m  
absorption seen in $\alpha$ Ori (Fig.1a) can be explained this way.
Also, neither emission nor absorption will appear from the MOLsphere
for some intermediate  values of $ R_{in}$  and the mid-infrared region of
$\alpha$ Ori (Fig.2a) may be such a case. Returning to $\mu$ Cep,
the value of  $ R_{in} \approx 2 R_{*} \approx 1300 R_{\odot} $ we have
suggested should have important implications. Especially, this result suggests
that water is absent in the region between $ r \approx R_{*}$ and $2 R_{*}$,
and water of the MOLsphere may be formed anew somewhere at
about $ R_{*}$ above the photosphere rather than the region very close to
the photosphere.  This fact should be kept in mind in considering
the origin of water in the  MOLsphere.

\section{CONCLUDING REMARKS}

We found  distinct emission of water at 6 $\mu$m and 40 $\mu$m regions
in the ISO spectrum of $\mu$ Cep in addition to the absorption bands
at 1.4, 1.9 and 2.7 $\mu$m in the near infrared. We showed that all 
these observed features can be explained consistently by a  simple model 
of the optically thick molecular sphere with modest temperature (e.g. 1500K) 
at modest distance above the photosphere (e.g. about one stellar radius).
What we have proposed, however, is not a model of the usual meaning but
simply an intuitive interpretation on the observed data, since our ``model'' 
cannot yet be constructed based on the first principle. 
In fact, the presence of the optically thick molecular envelope, 
which we  referred to as the MOLsphere for simplicity, has never been
predicted by any theory of stellar atmospheres so far as we are aware.
Thus, despite the simplicity of the picture we propose, the physics
behind the apparently simple phenomenon remains obscure and
offers a challenging problem to the theory of stellar atmospheres.

Recently, presence of water has been shown in several cool stars by ISO 
(Sect.1), but  interpretation of the observed spectra is by no means clear.
On the other hand, we could present a simple but consistent interpretation 
for the case of $\mu$ Cep and this may be due to fortunate situations 
as follow. {\it First}, $\mu$ Cep is unique in that water could be
observed through the near- to mid-infrared (including the Stratoscope 
data) and not only in absorption but also in emission. 
{\it Second}, $\mu$ Cep has the MOLsphere which  has just developed to 
an adequate size. For comparison, $\alpha$ Ori 
may be the case in which the MOLsphere is not yet developed enough to give
clear observed features, or emission and absorption (including the 
photospheric contributions) may be nearly canceled. On the contrary, 
the cooler supergiants may have well developed MOLsphere, the observed 
features from which may be heavily saturated and more difficult to interpret. 
{\it Third}, the spectrum of $\mu$ Cep shows almost no photospheric water 
band, unlike the case of the cooler objects which show strong photospheric
water bands difficult to discriminate from those originating in the MOLsphere.
{\it Fourth}, the MOLsphere of $\mu$ Cep  may be relatively free from 
complexities due to the dynamical effect such as by a large amplitude 
pulsation. For all these reasons, the spectra of $\mu$ Cep could be
interpreted relatively well and this may  provide the proto-type of 
the  MOLsphere, which should most probably exist in
the outer atmosphere of cool luminous stars in general.

We  finally confirmed that water exists not only in very cool luminous 
variables, but also in the early M supergiants and possibly in most M 
(super)giant stars. This fact implies that the presence of water is not due 
to some secondary effect such as a large amplitude pulsation but may 
be due to a more intrinsic property of cool stars. For example, the 
MOLsphere may have some connection with the chromosphere which also exists
in the outer atmosphere of the most cool stars, but its origin is 
by no means clear yet. By the opening of the infrared 
spectral region by ISO, we have now a  means by which to explore
the new component of the outer atmosphere which is rather warm (MOLsphere), 
in addition to the hot chromosphere and the cool expanding gas-dust envelope
known so far. The problem is how to understand these multiple components  
of different thermal and kinematical properties in a unified model, and 
we now have a possibility to arrive at a more balanced view of the stellar 
outer atmosphere by considering all its major constituents.

\acknowledgments
I thank W. Aoki, T. Tanab\'e, and I. Yamamura for their help and 
useful discussion on the ISO data.  This research is supported by 
the Grant-in-Aids for Scientific Research of the Ministry of Education, 
Culture, and Science No. 11640227.

\clearpage

\figcaption{
(a) Spectrum  of $\alpha$ Ori by the ISO SWS (resolution $R \approx 1600$)
and corrected for the interstellar reddening with $A_{v} = 0.5$ mag 
(Lee 1970). 
(b) The same but for $\mu$ Cep corrected for the interstellar reddening 
with $A_{v} = 1.5$ mag (Lee 1970).
(c) Absorption cross-section (cm$^{2}$ molecule$^{-1}$) of H$_{2}$O at 
$T =1500$K based on HITEMP (Rothman 1997). The black line is by high 
resolution ($R \approx 50000$) and the white line by 
low resolution ($R \approx 1600$).
\label{fig1}
}

\figcaption{
(a) Spectrum  of $\alpha$ Ori.   (b) Spectrum  of $\mu$ Cep. 
(c) Absorption cross-section of water at $T =1500$K. As for details,
see the legend to Fig.1.
\label{fig2}
}

\figcaption{
(a) The full spectrum  of $\mu$ Cep by the ISO SWS (corrected for the 
interstellar reddening) is shown by the red dots. 
The gaps around 4, 7, 12, 20, 29 $\mu$m are instrumental effect 
not yet corrected for. The predicted spectrum based on the MOLsphere 
(optically thick isothermal sphere of the water gas with $ T_{\rm ex} 
= 1500$K, $N_{\rm col} = 3.0 \times 10^{20}$ cm$^{-2}$, and located 
at $ R_{*} = 650 R_{\odot}$ above the photosphere) computed with the 
photospheric spectrum as the boundary condition is shown
by the green line (resolution $R \approx 50000$) and by the blue line
($R \approx 1600$), after being shifted vertically to fit
the observed spectrum. For comparison with observation, the effect of
dust emission should be added for $\lambda > 8 \mu$m.
The molecular bands indicated are originating in the photosphere
except for H$_{2}$O formed anew in the MOLsphere.
(b) A close-up of Fig.3a  around H$_{2}$O $\nu_{2}$ fundamentals.
 The observed spectrum  is shown by the red dots (identical with the ones 
in Fig.3a) and the predicted one  based on the MOLsphere by the blue line 
(identical with the one in Fig.3a).
(c) The region of the H$_{2}$O pure rotation lines in an expanded scale.
Now the dust emission is  added to the predicted spectrum from the 
MOLsphere (the blue line in Fig.3a), and the resulting spectrum is shown 
by the black line for comparison with the observed one (the red dots). 
\label{fig3}
}

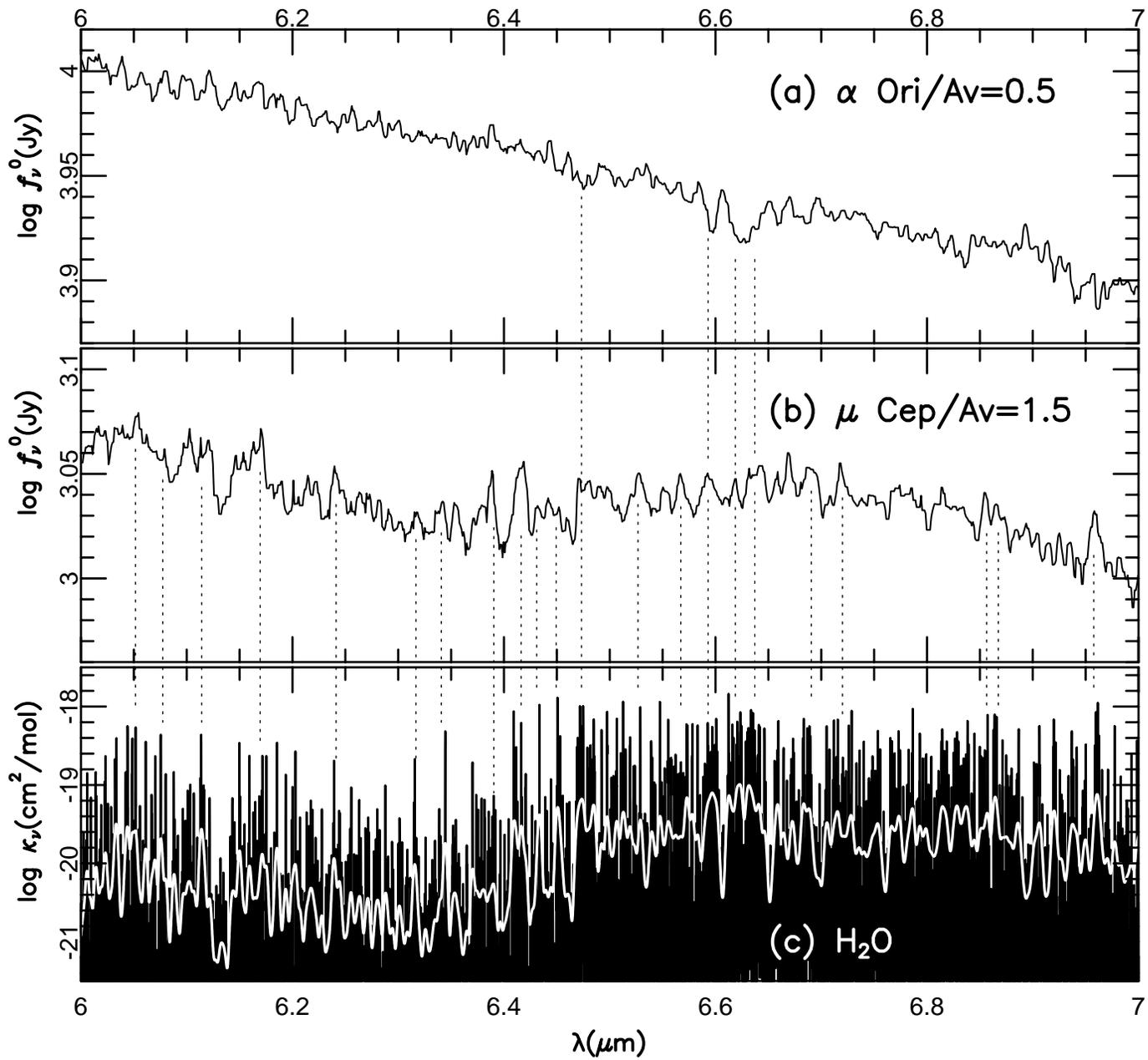
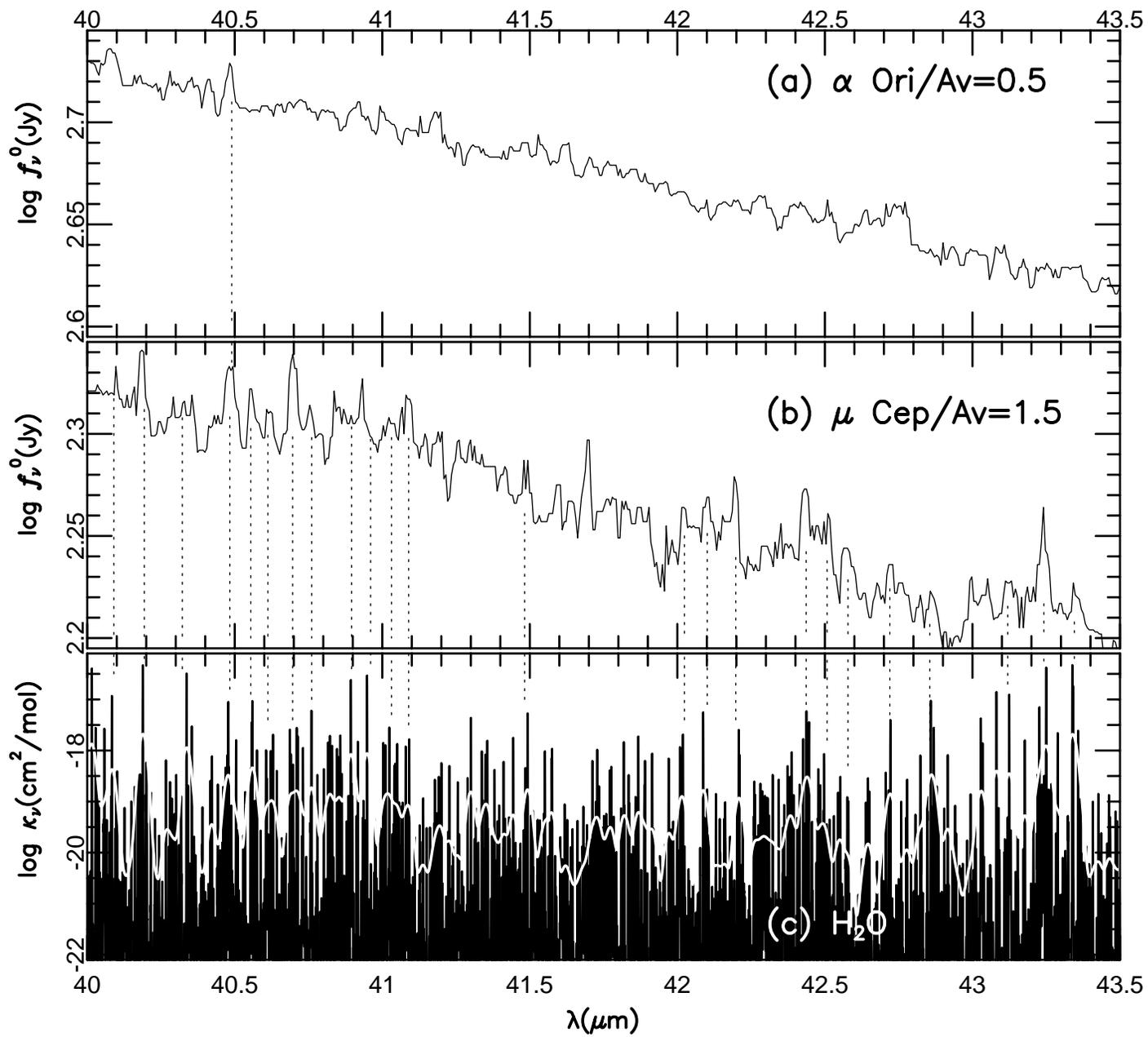
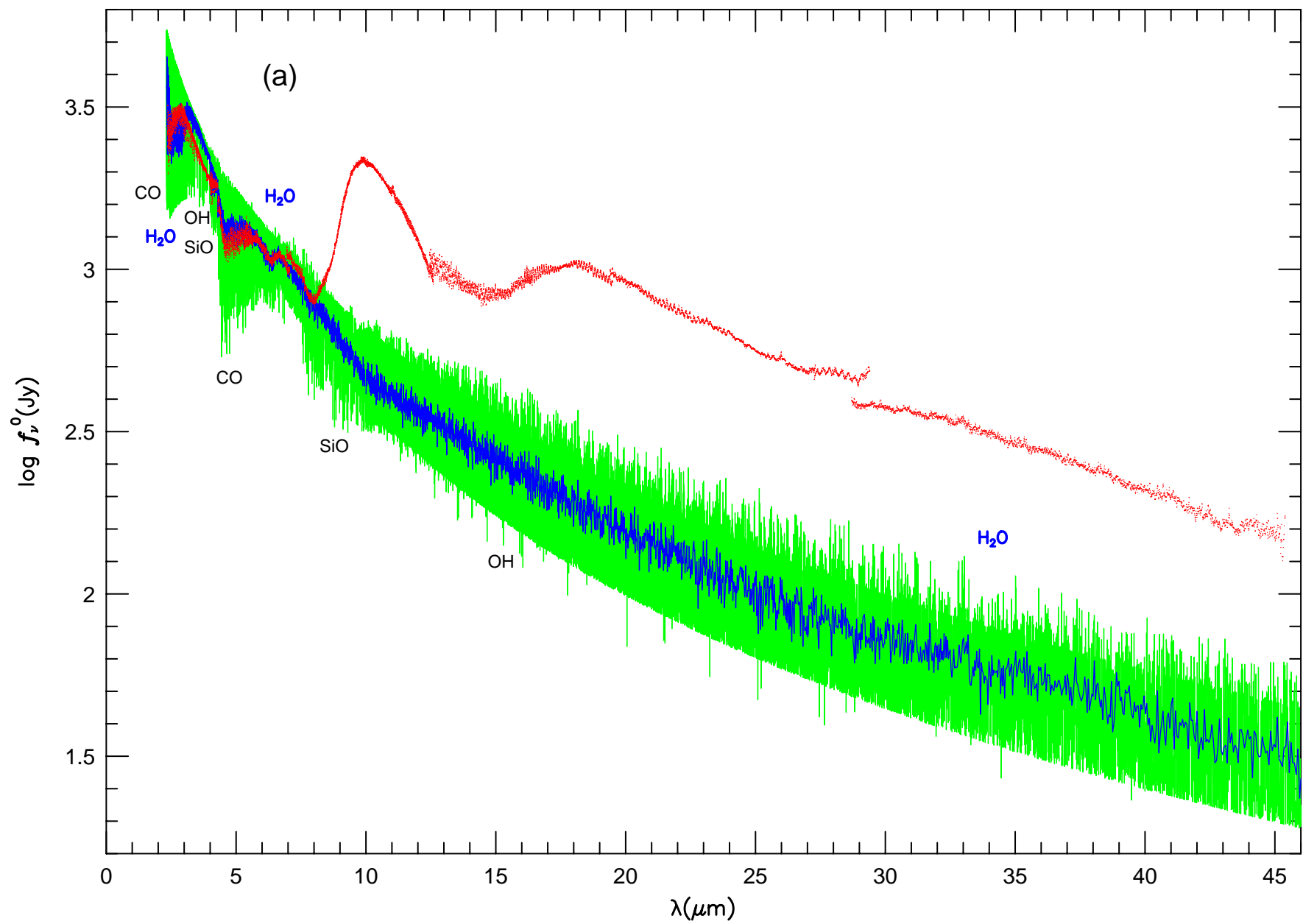
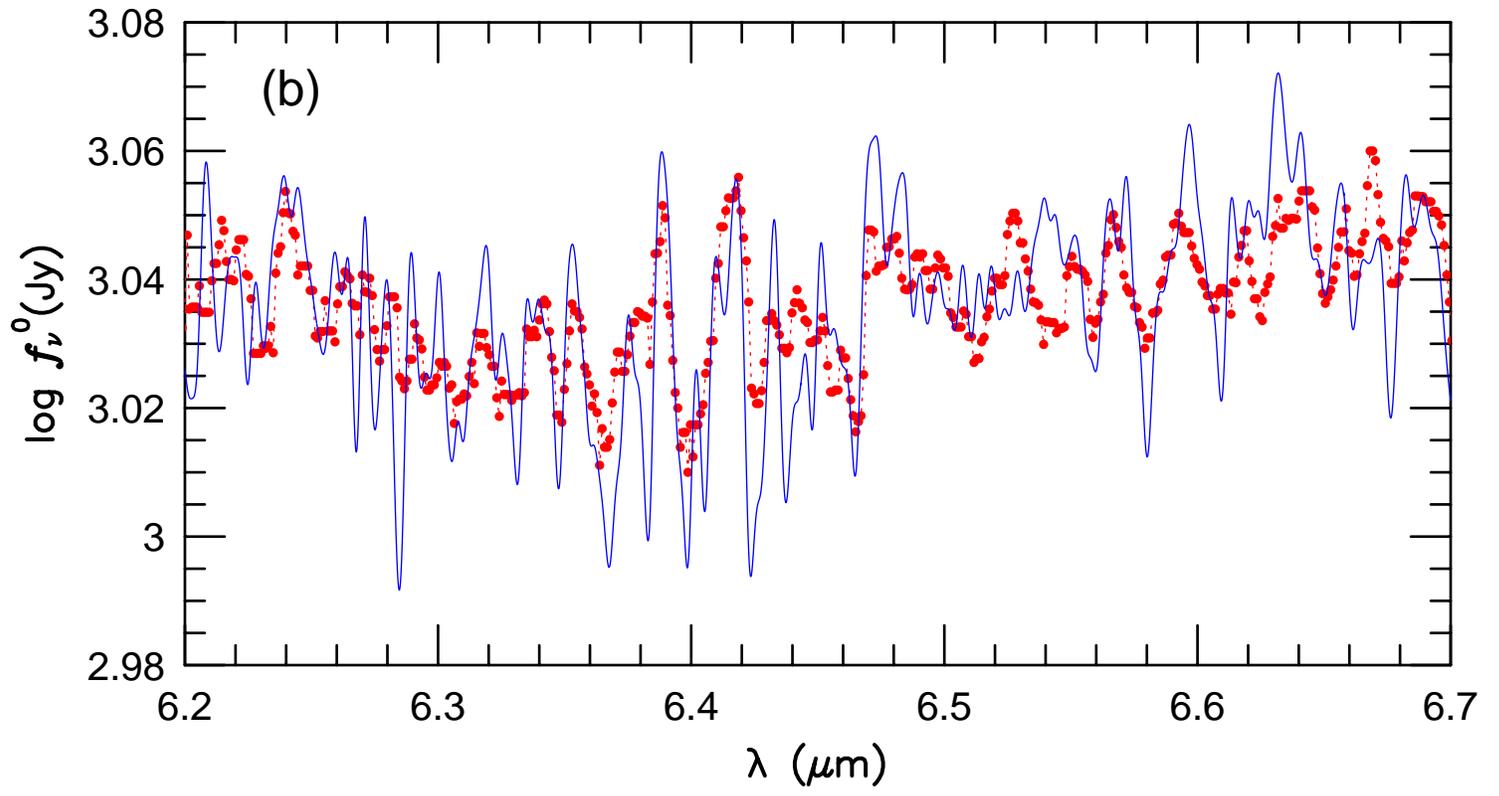
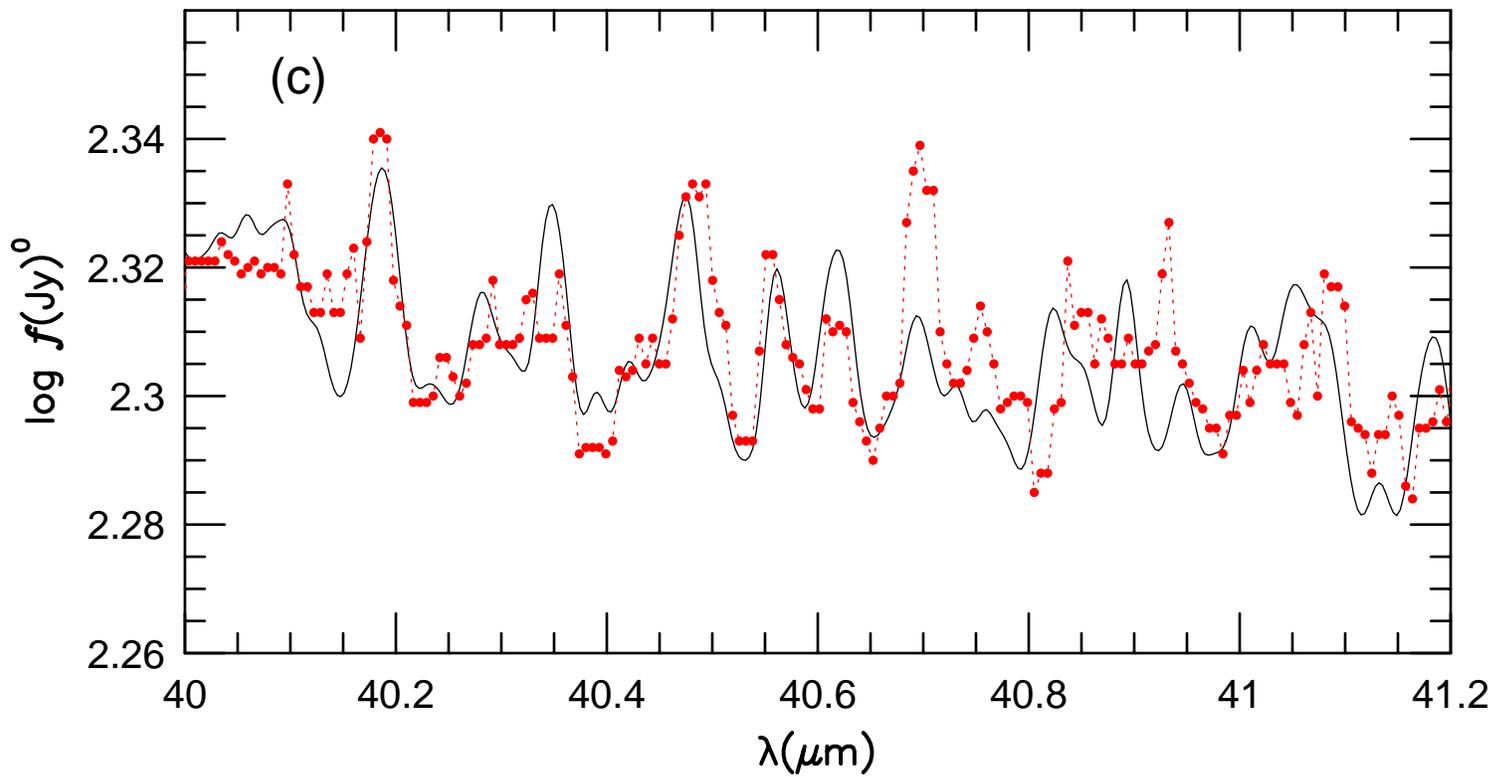
\end{document}